\documentclass{elsart}
\usepackage {lscape}
\usepackage{amssymb}
\usepackage{graphicx}
\usepackage{color}
\usepackage{rotating}
\begin{document}

\begin{frontmatter}
%\begin{fmffile}{diagrams}
\title{Shell model study of high-spin states and band terminations in $^{67}$As}

%\begin{vquote}
\author[1]{Vikas Kumar}
%\corref{cor1}\fnref{fn1}}
\ead{vikasphysicsiitr@gmail.com}
\author[2]{and Praveen C. Srivastava}
%\fnref{fn2}}
\ead{praveen.srivastava@ph.iitr.ac.in}
%\end{vquote}
%\begin{vquote}
 \address[1]{Department of Physics, Central University of Kashmir, Ganderbal 191 201, 
India}
 \address[2]{Department of Physics, Indian Institute of Technology, Roorkee 247 667, 
India}

\date{\today}
%%%%%%%%%%%%%%%%%%%%%%%%%%%%%%%%%%%%%%%%%%%%%%%%%%%%%%%%%%%
\begin{abstract}  
In the present work, recently available experimental data for different bands of $^{67}$As [Phys. Rev. C {\bf 98}, 024313 (2018)] have been interpreted within the framework of the shell model  in full $f_{5/2}pg_{9/2}$ model space using  JUN45 and jj44b effective interactions. 
The variation of $E$ - $E_{rot}$ energy versus spin for different bands
is shown to obtain useful information about band termination.
We have also reported  the electromagnetic transition probabilities, quadrupole and magnetic moments of $^{67}$As. 
Except for some tentative high-spin states, the results are in good agreement with the available experimental data.
%%The results are in good agreement with the available experimental data.
%%The wave functions, particularly, the specific proton and neutron configurations that are involved to generate the total angular momentum are discussed for selected energy states of $^{67}$As.
 The states $29/2^+$-$45/2^+$ are described by three particles in $g_{9/2}$, while $47/2^+$ and $51/2^+$ states are described by five particles in $g_{9/2}$. 
The shell model results corresponding to nine tentative states such as $41/2_1^+$ and $45/2_1^+$ in band-1b; $41/2_2^+$ and $45/2_2^+$ in band-1a; $43/2_2^+$, $47/2_2^+$  and $51/2_1^+$  in band-2a; 
$43/2_1^+$ and $47/2_1^+$ in band-2b are reported and discussed. 

\end{abstract}


\end{frontmatter}
% main text

%%%%%%%%%%%%%%%%%%%%%%%%%%%%%%%%%%%%%%%%%%%%%%%%%%%%%%%%%%%
\section{Introduction}

\label{intro}
\label{intro}
The band termination and nontermination is important properties of nuclei in the $Z, N = 28-50$ region. 
When we have few valence nucleons outside the $^{56}$Ni core then nuclei exhibit band termination, while if
valence nucleons are large then we have band nontermination. The experimental observation of band termination in non collective states of $^{62}$Zn (6 valence nucleons) is reported in Ref. \cite{62zn}. The evidence for nontermination of rotational bands is observed in $^{74}$Kr (18 valence nucleons) \cite{74kr}. The nontermination of yrast bands at maximum configuration spin in $^{73}$Kr is reported in Ref. \cite{teinhardt}. Thus, the nontermination is important phenomenon for nuclei approaching the middle of the $Z, N = 28-50$ region \cite{teinhardt}. The experimental evidence of band termination in $^{73}$Br is reported in Ref.\cite{73br}. The band termination in $^{69}$As is reported in Ref. \cite{69as}.  

In the more recent study of $^{67}$As having 11 nucleons outside $^{56}$Ni core exhibit band terminating structure \cite{R.Wadsworth}. The energy levels and $\gamma$-ray decay scheme of $^{67}$As have been studied in Ref. \cite{R.Wadsworth} using the $^{40}$Ca($^{36}$Ar, 2$\alpha$p)$^{67}$As reaction at Argonne National Laboratory. In this work, two new band structures have been reported and further, these two bands are connected to the previously known energy levels \cite{R. Orlandi}. The comparison of the experimental results for these two bands with theory using cranked Nilsson-Strutinsky model  is  reported in Ref. \cite{R.Wadsworth}, both theory and experiments suggest that these structures can be interpreted in terms of configurations that involve three $g_{9/2}$ particles \cite{R.Wadsworth}. 

The study of these newly populated terminating bands in $^{67}$As \cite{R.Wadsworth} using state-of-art large scale  shell model calculations is very important.
In this work, we have done  a comprehensive shell model study corresponding to different bands within $f_{5/2}pg_{9/2}$ model space using two different jj44b and JUN45 interactions. Also, we have reported the electromagnetic transition probabilities, quadrupole and magnetic moments.  In this region, the rotational features of $N=Z$ nuclei that possess large triaxiality  because of shell effect are reported in Ref. \cite{srivastava2}.
%%For high-spin states recently we have done extensive shell model calculations in this region \cite{srivastava1,gajpg,geproceedings,seeven,as,62ga,kumar,navin,palit1,palit2}. Thus shell model study of $^{67}$As using these effective interactions is very crucial.

             This paper is organized as follows: details about shell model calculations are given in section \ref{smc}. In section \ref{rd}, results and discussions are discussed. 
%%The decomposition of  the total wave function is discussed in section \ref{wd}.
 In section \ref{em}, the electromagnetic properties are discussed. Finally, summery is drawn in section \ref{sm}.

%%%%%%%%%%%%%%%%%%%%%%%%%%%%%%%%%%%%%%%%%%%%%%%%%%%%%%%%%%%
\section{Shell Model Hamiltonian} \label{smc}
In the present shell model calculations, $^{56}$Ni is taken as an inert core with  the spherical orbits $1p_{3/2}$, $0f_{5/2}$, $1p_{1/2}$ and $0g_{9/2}$ in the model space for both neutrons and protons. For calculations,  we used JUN45 and jj44b interactions. Brown and Lisetskiy developed the jj44b interaction \cite{jj44b} and further this interaction was fitted with 600 experimental binding energies and excitation energies of nuclei with $Z = 28-30$ and $N = 48-50$ available in this region. Here, 30 linear combinations of $JT$ coupled two-body matrix elements (TBME) are varied giving the rms deviation of about 250 keV from  experimental data. The single particle energies (spe)  are taken to be  -9.6566, -9.2859, -8.2695 and -5.8944 MeV for the 1$p_{3/2}$, 0$f_{5/2}$, 1$p_{1/2}$ and 0$g_{9/2}$ orbits respectively \cite{jj44b}.  The JUN45 \cite{jun45} interaction is based on Bonn-C potential, the  single-particle energies and two-body matrix elements were modified empirically with A = 63$\sim$69 mass region. The single-particle energies for the 1$p_{3/2}$, 0$f_{5/2}$, 1$p_{1/2}$ and 0$g_{9/2}$ orbitals are -9.828, -8.709, -7.839, and -6.262 MeV, respectively. 
The shell model calculations  were performed using the shell model codes  NuShellX \cite{NuShellX}, Antoine \cite{Antoine} and KSHELL \cite{KSHELL}.
%%%%%%%%%%%%%%%%%%%%%%%%%%%%%%%%%%%%%%%%%%%%%%%%%%%%%%%%%%%%
\section{Results and Discussions}
\label{rd}
\begin{figure}
\begin{center}
\includegraphics[width=10.0cm,height=8.0cm,clip]{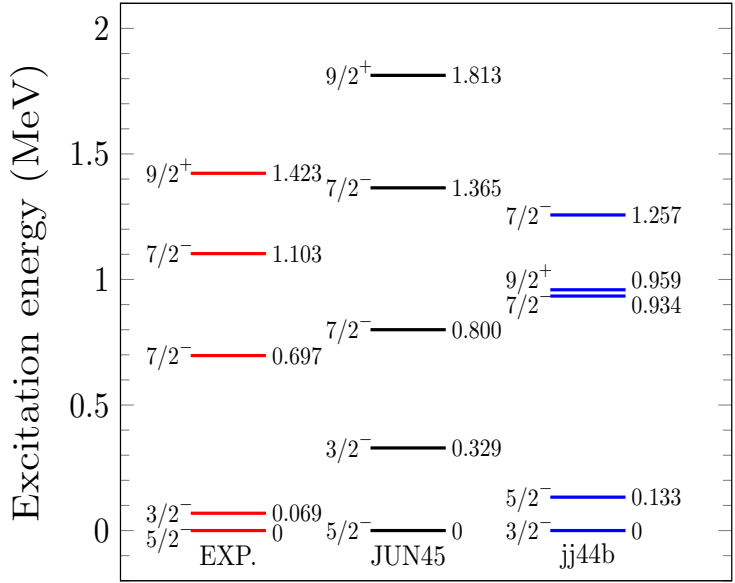}
\end{center}
\caption{Energy spectra of diffrent states bellow $9/2^+$ band head.}  
\label{below92}
\end{figure}

%%The comparison of calculated and experimental spectra for different bands of $^{67}$As nucleus  is given in Fig. \ref{bandsmjun45} and Fig. \ref{bandsmjj44b} using JUN45 and jj44b effective interactions, respectively.
In this section we have reported comprehensive shell model results of different bands corresponding to experimental data. 
We have calculated many  eigenvalues corresponding to each $J^{\pi}$  and then we have classified into different bands on the basis of dominant $E2$ transitions between them. In order to identify the band structures, we have connected the states with  a strong transition matrix elements between them and with similar configuration in the wave functions.
%% The occupancies of the protons and neutrons in the orbits for the levels in various bands are given in Fig 4, and Fig. 5  corresponding to  JUN45 and jj44b effective interactions, respectively. 
In the Fig \ref{below92}, we have shown comparison between calculated and experimental states below $9/2^+$. The JUN45 interaction correctly reproduced ground state and sequence of $5/2_1^-$-$3/2_1^-$-$7/2_1^-$-$7/2_2^-$-$9/2_1^+$ states.

\begin{figure}
\begin{center}
\includegraphics[width=14.5cm,height=13.5cm,clip]{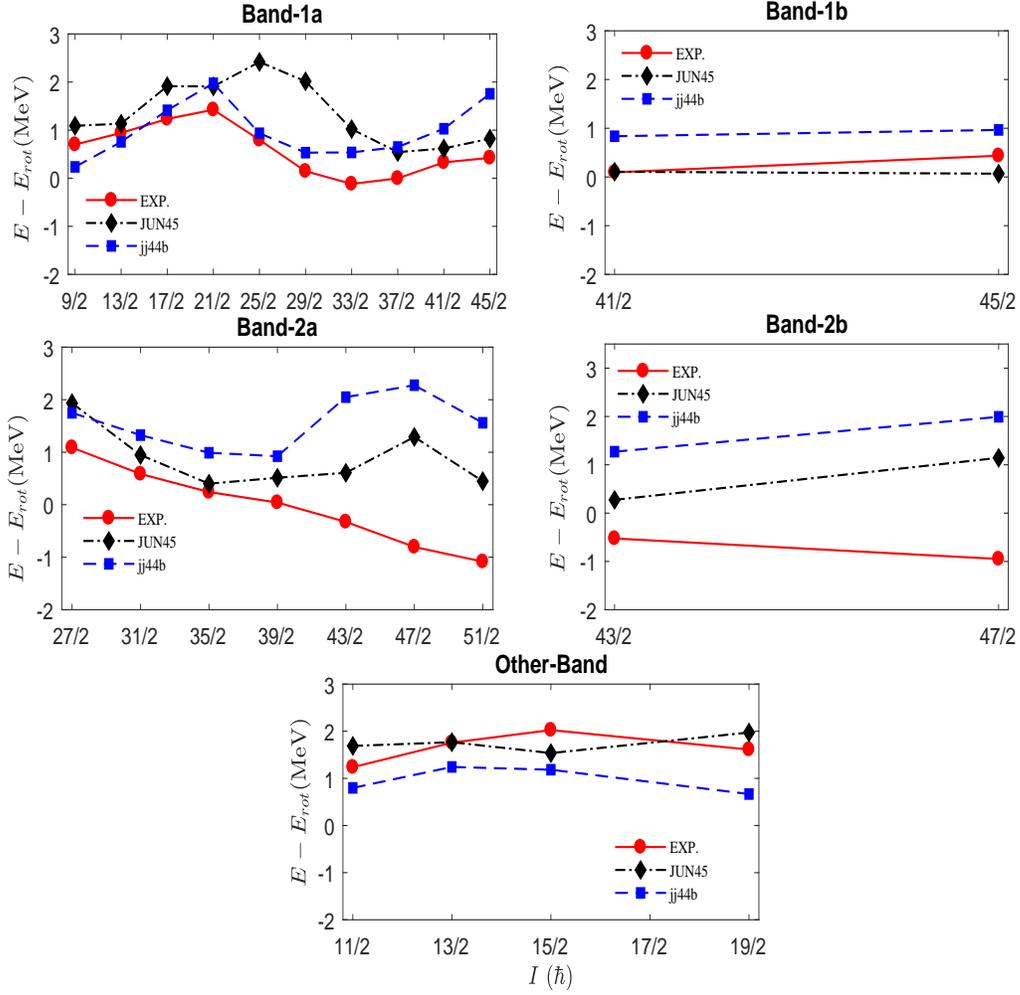}
\end{center}
\caption{ Calculated and experimental energies corresponding to different bands in $^{67}$As. A rotational reference $E_{rot}= 0.0292I(I+1)$ has been subtracted.}  
\label{bandref}
\end{figure}

The calculated and experimental energies for the different bands are shown in Fig. \ref{bandref}.  For convenience a rotational reference has been subtracted. It is seen how the rotational behavior and band termination, are reasonably  described by the shell model calculations.
The related energies of the band members as expressed by $E-E_{rot}$ form a specific configuration dependent curve as a function of spin.  The value of $E-E_{rot}$ become a minimum followed by a sharp increase before terminates. The kink in the curve is observed because of configuration changes. 
 The JUN45 interaction is showing band termination more reasonably for band-1a and band-2a in comparison to jj44b interaction.
It is interesting to note that, in the {band-2a}, beyond $39/2^+$, the experimentally observed intensity is reduced. This suggests structure change beyond this point. The shell model
result in Fig. \ref{bandref} is also showing different behaviour of  $E-E_{rot}$ curve after this point. 
In CNS calculation as reported in Ref. \cite{R.Wadsworth}, the $39/2^+$ is associated with a non-collective terminating configuration.

%% The JUN45 interaction is showing similar trend as in the experimental data for band-1a.  

 The shell model results corresponding to JUN45 and jj44b interactions are shown in Fig. \ref{bandsmjun45}. To identify the band structure, we have connected states with strong transitions matrix elements by lines as shown in Fig. \ref{bandsm3}. Also, these states are connected with similar dominant configuration in the wave functions.

\begin{figure}
\begin{center}
\hspace{-1cm}
\includegraphics[width=14.5cm,height=9.0cm,clip]{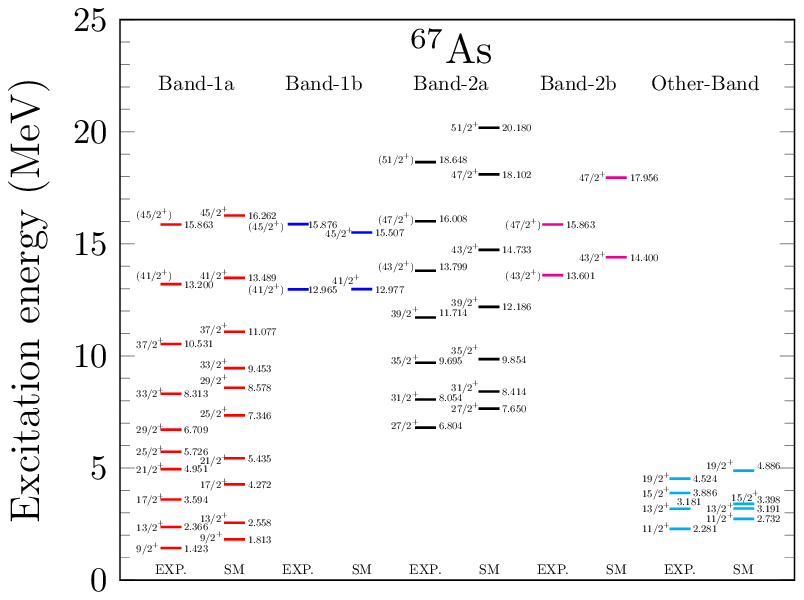}
%\includegraphics[width=8.5cm,height=7.0cm,clip]{67As_jj44b.eps}
%\caption{Comparison of shell model results with experimental data for different bands with JUN45 interaction. The band numbers in the figure are as
%per the convention used in the experimental paper \cite{R.Wadsworth}.}  
%\label{bandsmjun45} 
\end{center}
\end{figure}
\hspace{-2cm}
\begin{figure}
\begin{center}
\hspace{-1cm}
\includegraphics[width=14.5cm,height=9.0cm,clip]{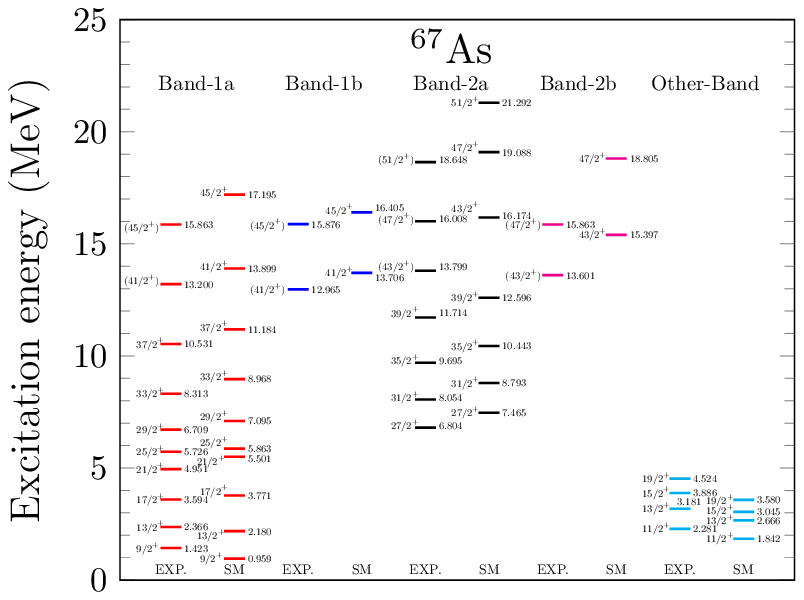}
\caption{Comparison of shell model results with experimental data for different bands with JUN45 and jj44b interactions. The band numbers in the figure are as
per the convention used in the experimental paper \cite{R.Wadsworth}.}  
\label{bandsmjun45} 
\end{center}
\end{figure}

%%%%%%%%%%%%%%%%%%%%%%%%%%%%%
\subsection{Band-1} 

In the band-1a the states $9/2^+$-$45/2^+$ are reported with JUN45 and jj44b interactions as shown in Fig. 2. The jj44b interaction results
are reasonable. The JUN45 interaction is showing similar trend as in the experimental data for band-1a, but for spins  $25/2^+$-$33/2^+$ results are not good.
%%Several new high-spin positive parity states up to the excitation energy 15.876 MeV are well produced by the shell model calculations in band-1a and band-1b.
%% All the sequences of the energy levels from both the shell model calculations are exactly matched with the experiments in this band. 
%%The calculated energy levels are slightly higher than the experiments this maybe because of the incompleteness of the basis state. 
 The energy difference between $33/2^+$~$\rightarrow$~$29/2^+$ transition is 1604 keV \cite{R.Wadsworth} in the experiment, whereas in the present work this energy differences are 875 and 1873 keV using JUN45 and jj44b interactions, respectively. 

 %%In band-1a, the first two states $9/2^+$ and $13/2^+$ are lower by 464 and 186 keV than experiment in jj44b, while 390 and 192 keV higher than experiment in JUN45. The positive parity energy states $17/2^+$, $21/2^+$, $25/2^+$, $29/2^+$, $33/2^+$, $37/2^+$, $41/2^+$, and $45/2^+$ are 678, 484, 1620, 1869, 1140, 546, 289, and 399 keV higher in JUN45; while 177, 550, 137, 386, 655, 653, 699, and 1332 keV higher in jj44b.
Both interactions predict $\pi$($p_{3/2}^{2}$,$f_{5/2}^{2}$,$g_{9/2}^{1}$)$\otimes \nu$($p_{3/2}^{2}$,$f_{5/2}^{2}$,$g_{9/2}^{2}$) configuration for $9/2^+$ and $13/2^+$ states.
%% In band-1b, the JUN45 prediction is very good for $41/2^+$ with a difference of only 12 keV with the experiment data, while this difference is 741 keV in jj44b. The $45/2^+$ state is lower by 369 keV in JUN45 while 529 keV higher in jj44b. 
There are two tentative spins $45/2_1^+$  and $45/2_2^+$ observed (band-1a and band-1b) in this experiment, we have reported results corresponding to these tentative spins from the shell model. A maximum spin of at least $49/2^+$ can  be achieved with contributions from three particles in $g_{9/2}$ orbit reported in Ref. \cite{R.Wadsworth} using configuration-dependent cranked Nilsson-Strutinsky (CNS) approach \cite{Afanasjev,Bengtsson,Carlsson} with Nilsson potential parameters \cite{Bengtsson}. The shell model configuration corresponding to positive-parity spin $45/2^+$  in band-1a and band-1b is $\pi(pf)_{6}^{4}(g_{9/2})_{4.5}^1$$\nu(pf)_{6}^{4}(g_{9/2})_{8}^2$, where $(pf)$ refers to the $p_{3/2}$, $f_{5/2}$ and $p_{1/2}$ orbits, the upper number represents the number of particles in the specified orbits and the lower number represents the maximum spin contribution from particles in these orbits, this configuration also suggests that three particles are required in $g_{9/2}$ orbital to achieve $I_{max}= 45/2^+$  spin in this band.  The single-particle energy of the $p_{1/2}$ orbit is lower in energy than $g_{9/2}$, but from  the nucleon occupation figures (Fig. \ref{occ_jun45} and Fig. \ref{occ_jj44b}) it is clear that $p_{1/2}$ orbit contributes little to the configuration.

%%%%%%%%%%%%%%%%%%%%%%%%%%
%\subsection{Band-1b}
\subsection{Band-2}

%% The spin sequence of the calculated positive parity energy levels in band-2a is the same as in the experiment. However, the energy levels $27/2^+$, $31/2^+$, $35/2^+$, $39/2^+$, $43/2^+$, $47/2^+$, and $51/2^+$ are 846, 360, 159, 472, 934, 2094, and 1532 keV higher in JUN45; while 661, 739, 748, 882, 2375, 3080, and 2644 keV higher in jj44b. In band-2b, the JUN45 results for the energy states $43/2^+$ and $47/2^+$ are 799 and 2093 keV higher; while 1796 and 2942 keV higher in jj44b. 
For band-2b, the results from both interaction are not good at high spin.
Overall the energy spectrum for JUN45 interaction is more closer to the experimental data up to $39/2^+$. 
In the band-2a, the $27/2^+$-$31/2^+$-$35/2^+$-$39/2^+$-$43/2^+$-$47/2^+$ states
are due to three $g_{9/2}$ particles, while in band-2b, $47/2^+$ state is due to five $g_{9/2}$ particles.

%The observed ground state spin and parity i.e., $5/2^-$  is also predicted by SM as a ground state with the configuration %$\pi(pf)_{4.5}^{5}(g_{9/2})_{0}^0$$\nu(pf)_{4}^{6}(g_{9/2})_{0}^0$. The SM successfully  produces the other observed %negative parity energy states (i.e., $3/2^-$  and $7/2^-$ ), and the configuration corresponding to these states are  the %same as the ground state. 

%%The tentative spin $({51/2}^+)$ observed in the experiment is also confirmed by the shell model calculations.
 The shell model result  supports the CNS prediction that such high-spin positive-parity states can only be formed in configuration involve five $g_{9/2}$ particles. The shell model configuration corresponding to positive-parity spin $51/2^+$ in band-2a is $\pi(pf)_{4.5}^{3}(g_{9/2})_{4}^2$$\nu(pf)_{4.5}^{3}(g_{9/2})_{8.5}^3$.  From the analysis of configuration, it is clear that only three $g_{9/2}$ particles are involved in the configuration for the positive-parity band at low energy. From the nucleon occupation figures (Fig. \ref{occ_jun45} and Fig. \ref{occ_jj44b}), the $p_{1/2}$ orbit has little contribution to the configuration, although its  single-particle energy is lower. 
 As reported in Ref. \cite{R.Wadsworth} that all transitions in band 2 are most likely not stretched $B(E2)$’s because for example the lowest $51/2^+$ state (with five $g_{9/2}$ particles) is calculated at a much higher energy. This is supported by our shell model calculations here the difference between calculations and experiment becomes much larger for the $47/2^+$ and $51/2^+$  states in band 2.

%%As shown in Fig. \ref{bandsm5}, for identifying band structures, we have connected by lines the states with strong transition %%matrix elements between them and with similar dominant configuration in the wave functions.

\begin{figure}
\begin{center}
\includegraphics[width=6.5cm,height=7.0cm,clip]{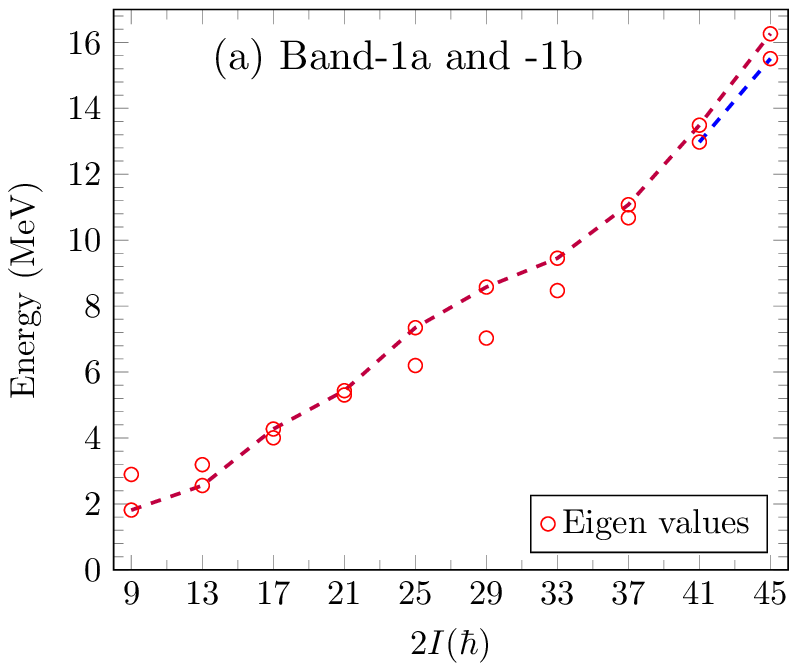}
\includegraphics[width=6.5cm,height=7.2cm,clip]{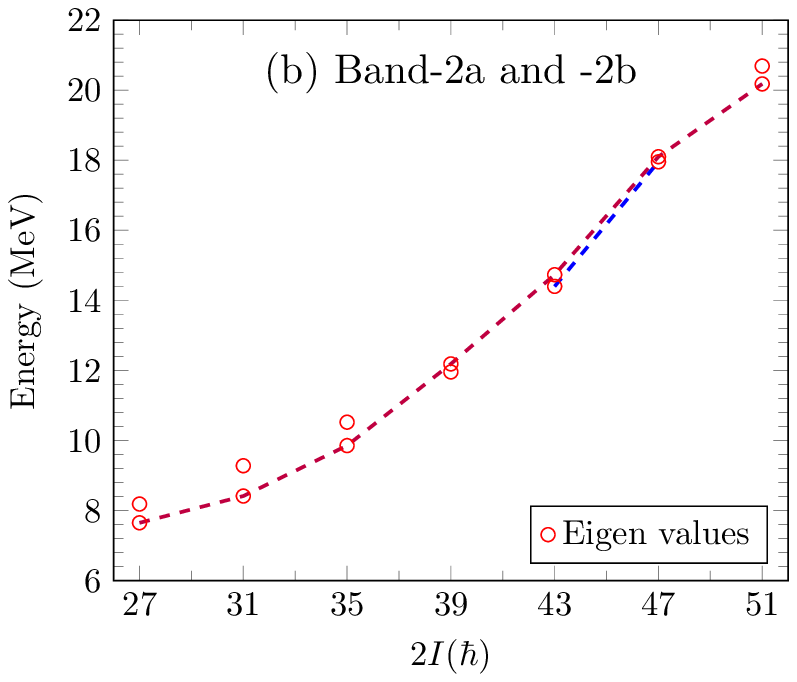}
\includegraphics[width=6.5cm,height=7.2cm,clip]{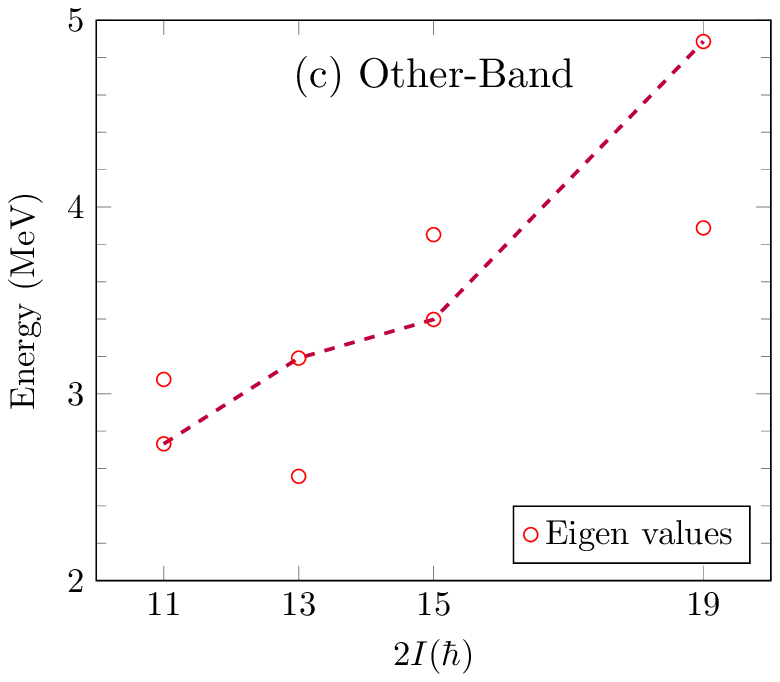}
\end{center}
\caption{Shell model predictions for different band in $^{67}$As with JUN45 interaction.}  
\label{bandsm3}
\end{figure}

%%%%%%%%%%%%%%%%%%%%%%%%%%%%%%%%%%%
\subsection{Other-Band}

 The positive parity energy states $11/2^+$, $13/2^+$, $15/2^+$, and $19/2^+$ at 2281, 3181, 3886, and 4524 keV in Fig. 1 of Ref. \cite{R.Wadsworth} is defined as other band in the present work.
The results of jj44b interaction shows similar trend as in the experiment.
 These energy levels are reasonably produced using both the shell model calculations.
 %%The sequence of the energy states in the experiment are exactly matched with the SM calculations.
 The configuration for different energy states in this band indicate that one particle is required in $g_{9/2}$ orbital to produce these spins and parity.
\begin{figure*}
\begin{center}
\resizebox{0.78\textwidth}{0.38\textwidth}{\includegraphics{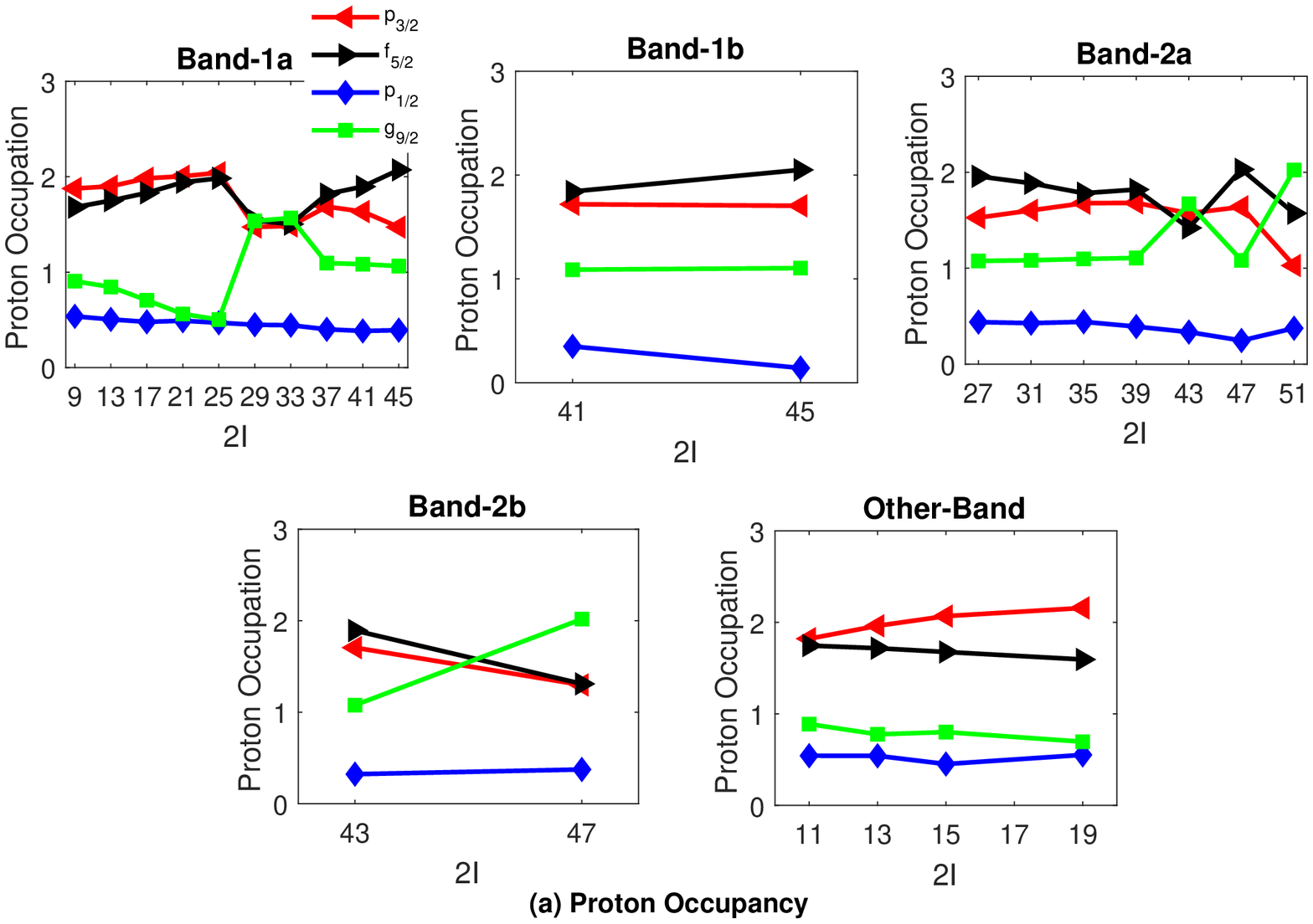}}
\resizebox{0.78\textwidth}{0.38\textwidth}{\includegraphics{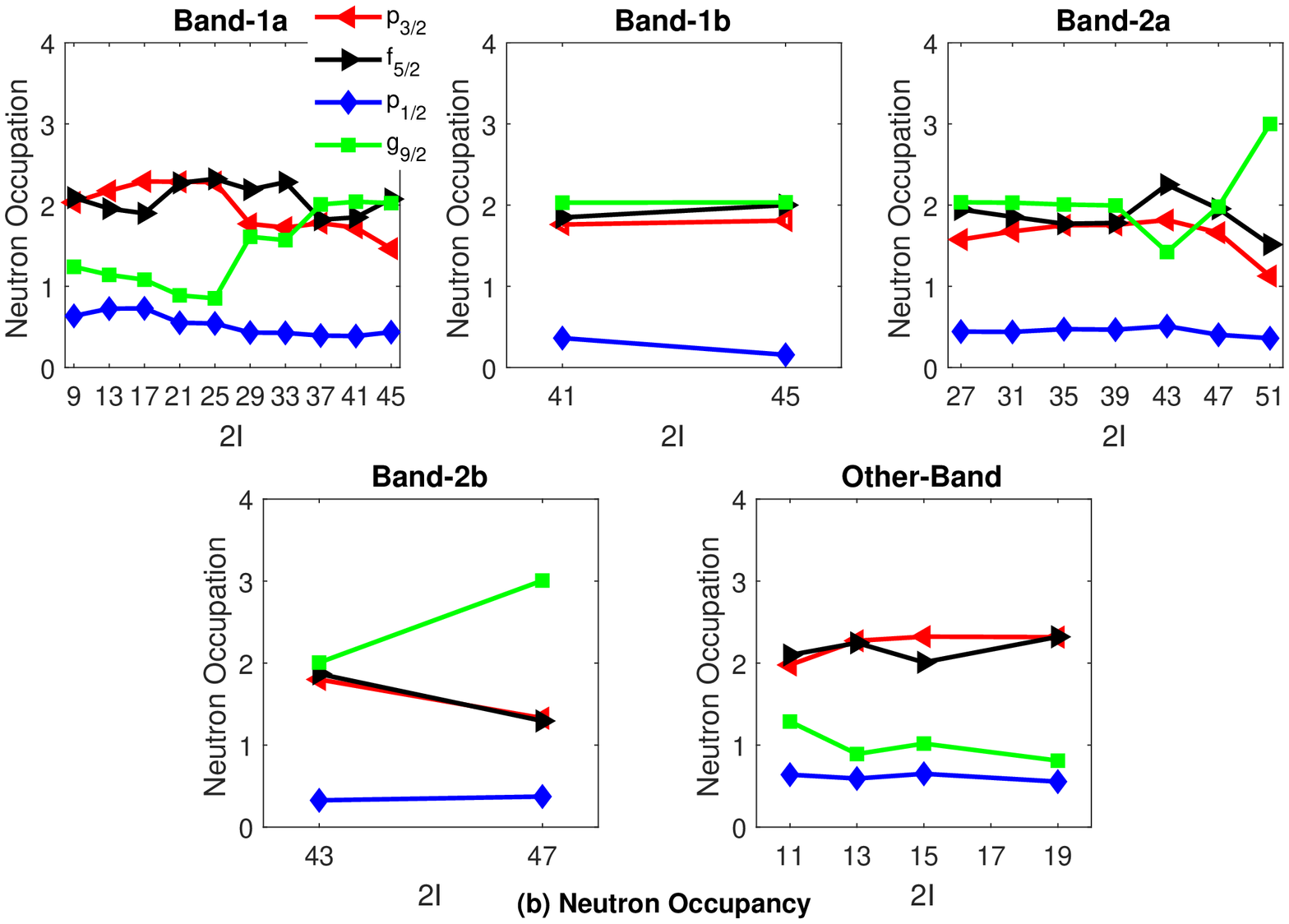}}
\caption{\label{occ_jun45} Occupancies of different states for $^{67}${As} using JUN45 interaction.}
\resizebox{0.78\textwidth}{0.38\textwidth}{\includegraphics{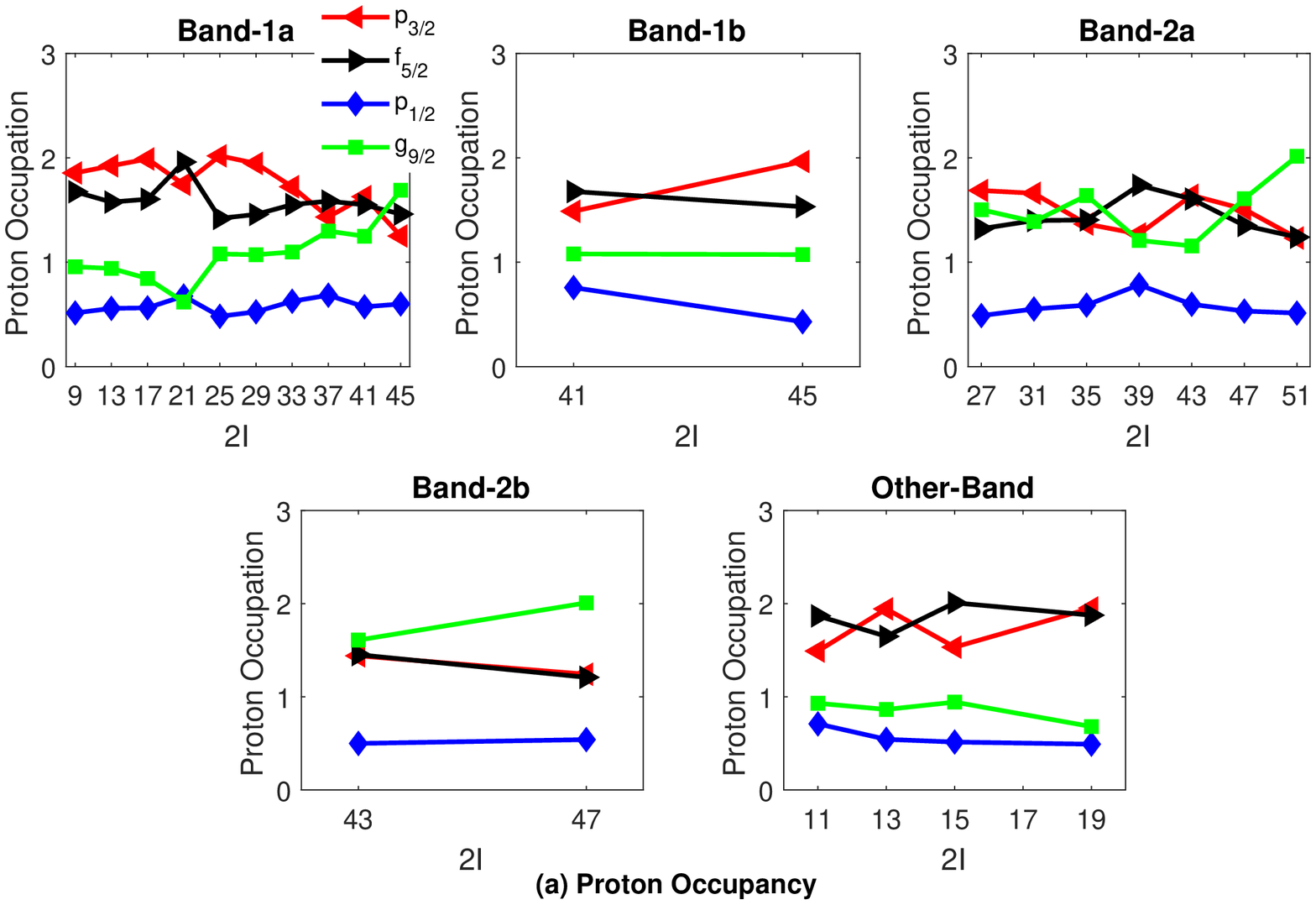}}
\resizebox{0.78\textwidth}{0.38\textwidth}{\includegraphics{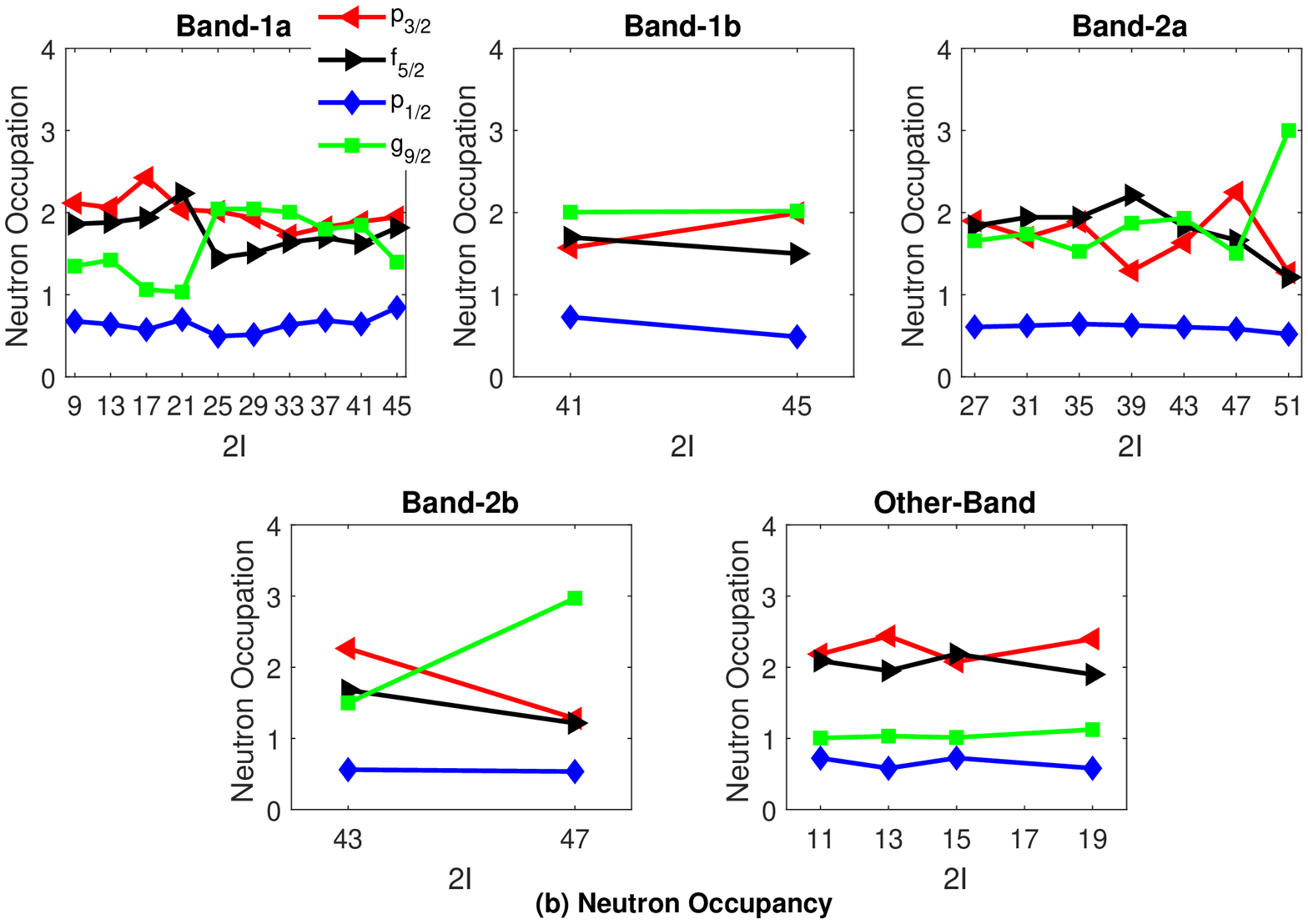}}
\caption{\label{occ_jun45} Occupancies of different states for $^{67}${As} using jj44b interaction.}
\end{center}  
\end{figure*}
%%%%%%%%%%%%%%%%%%%%%%%%%%%%%%%%%%%%%%%%%%%%%%%%%%%%%%%%%%%%%%
%%%%%%%%%%%%%%%%%%%%%%%%%%%%%%%%%%%%%%%%%%%%%%%%%%%%%%%%%%%%%%

%%%%%%%%%%%%%%%%%%%%%%%%%%%%%%%%%%%%%%%%%
\section{Electromagnetic Properties}
\label{em}

In Table \ref{table3} and \ref{table4}, we have reported calculated $B(E2)$ values for different transitions in different bands using JUN45 and jj44b effective interactions, respectively. 
The calculated bands are well connected with larger $B(E2)$ values. However, in our calculation for band-1a with JUN45, the $B(E2;29/2^{+}_{2} \rightarrow 25/2^{+}_{2}$) is very small. This is because there is a drastic change in the $g_{9/2}$ occupancy between $29/2^{+}_{2}$ and $25/2^{+}_{2}$ levels.  Shell model predicts the structure of the $29/2^{+}_{2}$ level to be quite different from that of the other levels lying below. Similarly, in the case of jj44b, for band-1a, the $B(E2;25/2^{+}_{2} \rightarrow 21/2^{+}_{2}$) is very small. This is because there is a drastic change in the $g_{9/2}$ occupancy between $25/2^{+}_{2}$ and $21/2^{+}_{2}$ levels. In the case of band-2, the $B(E2;51/2^{+}_{1} \rightarrow 47/2^{+}_{2})$ is very small, this is because of configuration changes, to generate  $51/2^{+}_{1}$ state we need three neutrons in $g_{9/2}$, while only two neutrons in $g_{9/2}$ for $47/2^{+}_{2}$. To know more about the $B(E2)$ values corresponding to states not belong to band with JUN45 interaction is shown in Table \ref{table6}.
For the calculation of magnetic moments in the present work, we have used $g_s^{\rm eff}$ = $g_s^{\rm free}$. The results of electric quadrupole and magnetic moments are listed in Table \ref{table5} for different bands using JUN45 and jj44b effective interactions. The calculated results from both the calculations are in a good agreement with each other.
These predicted results might be very useful to compare upcoming experimental data.

%%%%%%%%%%%%%%%%%%%%%%%%%%%%%%%

\begin{table}
\begin{center}
\caption{ Calculated $B(E2)$ values (in $e^2fm^4$) for $^{67}$As using JUN45 interaction with  $e_{\pi} =1.5e$ ; $e_{\nu} =0.5e$.}
\resizebox{14.0cm}{!}{
\begin{tabular}{cc||ccc }
\hline
 ~~Transition   &~~ Band -1a  ~&  ~~Transition  & Band -2a   \\    
\hline

$B(E2;13/2^{+}_{1} \rightarrow 9/2^{+}_{1}$)\hspace{0.5cm} &  177  & $B(E2;31/2^{+}_{1} \rightarrow 27/2^{+}_{1}$)\hspace{0.5cm} &  293   \\
$B(E2;17/2^{+}_{2} \rightarrow 13/2^{+}_{1}$)\hspace{0.5cm} & 145   & $B(E2;35/2^{+}_{1} \rightarrow 31/2^{+}_{1}$)\hspace{0.5cm} &  257 \\ 
$B(E2;21/2^{+}_{2} \rightarrow 17/2^{+}_{2}$)\hspace{0.5cm} & 26  & $B(E2;39/2^{+}_{2} \rightarrow 35/2^{+}_{1}$)\hspace{0.5cm} &  118    \\ 
$B(E2;25/2^{+}_{2} \rightarrow 21/2^{+}_{2}$)\hspace{0.5cm} &  277 & $B(E2;43/2^{+}_{2} \rightarrow 39/2^{+}_{2}$)\hspace{0.5cm} &  24  \\ 
$B(E2;29/2^{+}_{2} \rightarrow 25/2^{+}_{2}$)\hspace{0.5cm}  & 0.12 & $B(E2;47/2^{+}_{2} \rightarrow 43/2^{+}_{2}$)\hspace{0.5cm} &  3     \\ 
$B(E2;33/2^{+}_{2} \rightarrow 29/2^{+}_{2}$)\hspace{0.5cm}  &  295 & $B(E2;51/2^{+}_{1} \rightarrow 47/2^{+}_{2}$)\hspace{0.5cm} &  0.02  \\ 
$B(E2;37/2^{+}_{2} \rightarrow 33/2^{+}_{2}$)\hspace{0.5cm}  & 79 &  &     \\ 
$B(E2;41/2^{+}_{2} \rightarrow 37/2^{+}_{2}$)\hspace{0.5cm}  & 164  &  &    \\ 
$B(E2;45/2^{+}_{2} \rightarrow 41/2^{+}_{2}$)\hspace{0.5cm}  & 23  &  &    \\ 
\hline
 ~~Transition   &~~ Band -1b  ~&  ~~Transition  & Band -2b   \\    
\hline

$B(E2;45/2^{+}_{1} \rightarrow 41/2^{+}_{1}$)\hspace{0.5cm} &  113  & $B(E2;47/2^{+}_{1} \rightarrow 43/2^{+}_{1}$)\hspace{0.5cm} &  0.002  \\
\hline
 ~~Transition   &~~ Other Band  ~&  ~~  &   \\    
\hline

$B(E2;13/2^{+}_{2} \rightarrow 11/2^{+}_{1}$)\hspace{0.5cm} &  90  &  &    \\
$B(E2;15/2^{+}_{1} \rightarrow 13/2^{+}_{2}$)\hspace{0.5cm} & 37   &  &   \\ 
$B(E2;19/2^{+}_{2} \rightarrow 15/2^{+}_{1}$)\hspace{0.5cm} & 169   &  &   \\ 
\hline        
\end{tabular}
\label{table3}}
\end{center}
\end{table}
%%%%%%%%%%%%%%%%%%%%%%%%%%%%%%%%%%%%%%%%%%%%%%%%%%%%%%%%%%%%%%%%%%%%%%%%%%%%%%
\begin{table}
\begin{center}
\caption{ Calculated $B(E2)$ values (in $e^2fm^4$) for $^{67}$As using jj44b interaction with  $e_{\pi} =1.5e$ ; $e_{\nu} =0.5e$.}
\resizebox{14.0cm}{!}{
\begin{tabular}{cc||ccc }
\hline
 ~~Transition   &~~ Band -1a  ~&  ~~Transition  & Band -2a   \\    
\hline

$B(E2;13/2^{+}_{1} \rightarrow 9/2^{+}_{1}$)\hspace{0.5cm} &  237  & $B(E2;31/2^{+}_{1} \rightarrow 27/2^{+}_{1}$)\hspace{0.5cm} &  236   \\
$B(E2;17/2^{+}_{2} \rightarrow 13/2^{+}_{1}$)\hspace{0.5cm}&167    & $B(E2;35/2^{+}_{1} \rightarrow 31/2^{+}_{1}$)\hspace{0.5cm} &  104   \\ 
$B(E2;21/2^{+}_{2} \rightarrow 17/2^{+}_{2}$)\hspace{0.5cm}& 198   & $B(E2;39/2^{+}_{2} \rightarrow 35/2^{+}_{1}$)\hspace{0.5cm} &  8     \\ 
$B(E2;25/2^{+}_{1} \rightarrow 21/2^{+}_{2}$)\hspace{0.5cm}&  2.48 & $B(E2;43/2^{+}_{2} \rightarrow 39/2^{+}_{2}$)\hspace{0.5cm} &  20    \\ 
$B(E2;29/2^{+}_{1} \rightarrow 25/2^{+}_{1}$)\hspace{0.5cm}& 309   & $B(E2;47/2^{+}_{2} \rightarrow 43/2^{+}_{2}$)\hspace{0.5cm} & 3      \\ 
$B(E2;33/2^{+}_{1} \rightarrow 29/2^{+}_{1}$)\hspace{0.5cm}&  250  & $B(E2;51/2^{+}_{1} \rightarrow 47/2^{+}_{2}$)\hspace{0.5cm} &  5     \\ 
$B(E2;37/2^{+}_{1} \rightarrow 33/2^{+}_{1}$)\hspace{0.5cm}& 170   &  &     \\ 
$B(E2;41/2^{+}_{2} \rightarrow 37/2^{+}_{1}$)\hspace{0.5cm}& 83   &  &    \\ 
$B(E2;45/2^{+}_{2} \rightarrow 41/2^{+}_{2}$)\hspace{0.5cm}& 43    &  &    \\ 
\hline
 ~~Transition   &~~ Band -1b  ~&  ~~Transition  & Band -2b   \\    
\hline

$B(E2;45/2^{+}_{1} \rightarrow 41/2^{+}_{1}$)\hspace{0.5cm} &  94  & $B(E2;47/2^{+}_{1} \rightarrow 43/2^{+}_{1}$)\hspace{0.5cm} &  8  \\
\hline
 ~~Transition   &~~ Other Band  ~&  ~~  &   \\    
\hline

$B(E2;13/2^{+}_{2} \rightarrow 11/2^{+}_{1}$)\hspace{0.5cm} &  115  &  &    \\
$B(E2;15/2^{+}_{1} \rightarrow 13/2^{+}_{2}$)\hspace{0.5cm} & 24   &  &   \\ 
$B(E2;19/2^{+}_{2} \rightarrow 15/2^{+}_{1}$)\hspace{0.5cm} & 77   &  &   \\ 
\hline 
\end{tabular}
\label{table4}}
\end{center}
\end{table}

%\vspace{6cm}

\begin{table}
\caption{The calculated $B(E2)$ values (in $e^2fm^4$ with $e_p$=1.5e, $e_n$=0.5e) for other transitions not belong to band with JUN45 interaction.}
%\resizebox{15.0cm}{11.0cm}{
%\label{tab:table3}
\hspace{-1.5cm}
\vspace{1.5cm}
\label{be2n}
\begin{tabular}{|c|c|c|c|c|c|c|c|c|c|c|}
\hline 
Transition ($J_i \rightarrow (J-2)_i)$ & $13/2^{+}_{2}$  & $17/2^{+}_{1}$   & $21/2^{+}_{1}$  & $25/2^{+}_{1}$   & $29/2^{+}_{1}$   & $33/2^{+}_{1}$  & $37/2^{+}_{1}$  & $41/2^{+}_{1}$   & $45/2^{+}_{1}$ \tabularnewline
\hline 
$B(E2)$              & 272  & 3 & 32  & $<$1  & 286 & 253 & 189 & 2  & 113 \tabularnewline
\hline 
\hline 
Transition ($J_i \rightarrow (J-2)_i)$ & $31/2^{+}_{2}$  & $35/2^{+}_{2}$   & $39/2^{+}_{2}$  & $43/2^{+}_{1}$   & $47/2^{+}_{1}$   & $51/2^{+}_{1}$  &  &   &  \tabularnewline
\hline 
$B(E2)$              & 1.42 & 118 & 9  & 52  & $<$ 1& 177 &  &  &  \tabularnewline
\hline 
\hline 
Transition ($J_i \rightarrow (J-2)_i)$& $47/2^{+}_{2}$  &    &  &   &    &  &  &   &  \tabularnewline
\hline 
$B(E2)$              & 3 &  &   &   &  &  &  &  &  \tabularnewline
\hline 
\hline 
Transition ($J_i \rightarrow (J-2)_i)$ & $13/2^{+}_{1}$  & $15/2^{+}_{1}$    & $19/2^{+}_{1}$  &   &    &  &  &   &  \tabularnewline
\hline 
$B(E2)$              & 8 & 185 & 2  &   &  &  &  &  &  \tabularnewline
\hline 
\hline 
\end{tabular}
\label{table6}
\end{table}
%%%%%%%%%%%%%%%%%%%%%%%%%%%%%%%%%%%%%%%%%%%%%%%%%%%%%%%%%%%%%%%%%%%%%%%%%%%%%%%%%%%%%%%%%%%

\begin{table}
\begin{center}
\caption{The calculated electric quadrupole moments $Q_s$ (in $eb$)  and 
magnetic moments  $\mu$ (in $\mu_N$). The effective charges $e_p$=1.5e, $e_n$=0.5e and  $g_s^{\rm eff}$ = $g_s^{\rm free}$ are used. No experimental data are available.}
\resizebox{15.0cm}{11.0cm}{
%\label{tab:table3}
\label{qm}
\begin{tabular}{c|c|c|c|c|c|c|c} 
\hline
\multicolumn{3}{c}{{\bf Band-1a}}                                         &   &\multicolumn{3}{c}{{\bf Band-2a} } \\
\hline
              &              & Electric Moments    &  Magnetic Moments                 &        &      & Electric Moments    &  Magnetic Moments    \\ 
\hline
   $9/2^+$    & JUN45        &  -0.642     &   +4.744   &  $27/2^+$   & JUN45        &  -0.755     &   +4.150      \\ 
              & jj44b        &  -0.667     &   +4.805      &        & jj44b        &  -0.943     &   +7.253     \\ 
\hline        
  $13/2^+$    & JUN45        &  -0.787     & +5.912     &  $31/2^+$     & JUN45        &  -0.809     & +5.100    \\
              & jj44b        &  -0.779     & +5.995         &         & jj44b        &  -0.954     & +7.885        \\       
\hline         
    $17/2^+$   & JUN45        &  -0.762     & +6.419   &   $35/2^+$   & JUN45        &  -0.840     &   +6.270      \\ 
              & jj44b        &  -0.594     & +5.096        &       & jj44b        &  -0.922     &   +10.111     \\ 
\hline        
$21/2^+$      & JUN45        &  -0.536     & +4.328      &      $39/2^+$   & JUN45        &  -0.685     & +7.254        \\
              & jj44b        &  -0.589      & +3.645         &         & jj44b        &  -0.877     & +10.610        \\ 
\hline        
$25/2^+$     & JUN45        &  -0.543     & +5.136          &        $43/2^+$  & JUN45        &  -0.844     &   +11.284      \\ 
              & jj44b        &  -0.975      & +3.694        &           & jj44b        &  -0.811     &   +8.450     \\                     
\hline         
$29/2^+$     & JUN45        &  -0.844    & +7.541    &    $47/2^+$         & JUN45        &  -0.620     & +9.566        \\
              & jj44b        & -1.016     & +4.625         &            & jj44b        &  -0.870     & +12.112        \\              
\hline        
$33/2^+$     & JUN45        & -0.879     & +8.528     &      $51/2^+$   & JUN45        &  -0.894     & +11.055        \\
              & jj44b        & -1.034      & +5.827         &       & jj44b        &  -0.978    & +10.929        \\ 
\hline        
              &             &               &                    &   \multicolumn{4}{c}{{\bf Band -2b}}  \\
              \hline
$37/2^+$     & JUN45        & -0.371     & +6.738     &    $43/2^+$    & JUN45        &  -0.632     &   +8.411      \\ 
              & jj44b        & -1.014      & +8.229         &         & jj44b        &  -0.867     &   +11.271     \\ 
\hline        
$41/2^+$     & JUN45        & -0.660     & +7.643      &      $47/2^+$    & JUN45        &  -0.944     & +9.886        \\
              & jj44b        & -0.860      & +8.665         &         & jj44b        &  -0.954     & +10.043        \\  
              \hline
 &             &               &                    &          \multicolumn{4}{c}{{\bf Other Band}} \\
 \hline
$45/2^+$     & JUN45        &  -0.733     & +8.385                &         $11/2^+$     & JUN45        &  -0.369     &   +4.773      \\ 
              & jj44b        &  -0.960      & +11.649        &       & jj44b        &  -0.647     &   +3.128     \\                       
\hline
\multicolumn{3}{c}{{\bf Band-1b}}  &   &   &   &  \\
\hline
$41/2^+$      & JUN45        &  -0.649     &   +7.628         &      $13/2^+$    & JUN45        &  -0.415     & +4.118        \\
              & jj44b        &  -0.943     &   +7.706      &           & jj44b        &  -0.450     & +3.973        \\    
\hline        
$45/2^+$    & JUN45        &  -0.572     & +8.985  &       $15/2^+$    & JUN45        &  -0.535     &   +5.600      \\ 
              & jj44b        &  -0.793     & +8.387         &         & jj44b        &  -0.708     &   +4.614     \\ 

              \hline
              
    &          &         &             &       $19/2^+$   & JUN45        &  -0.573     & +5.600        \\
              &         &       &         &           & jj44b        &  -0.523     & +3.034        \\               
\hline                                

\end{tabular}}
\label{table5}
\end{center}
\end{table}
%\footnotetext{No experimental values exist}

%%%%%%%%%%%%%%%%%%%%%%%%%%%%%%%%%%%%%%%%%%%%%%%%
%\newpage
\section{Summary}
\label{sm}
Motivated by recent experimental  data of different bands \cite{R.Wadsworth} for $^{67}$As, we have reported the comprehensive shell model study of different bands in $f_{5/2}pg_{9/2}$ model space using JUN45 and jj44b effective interactions. The following broad conclusions are drawn:

\begin{itemize}

\item The high spin structures of negative and positive parity bands are successfully described by both the effective interactions for the full $f_{5/2}pg_{9/2}$ model space.

%\item Some tentative spins-parity high-energy states \cite{R.Wadsworth}  such as $41/2_1^+$ and $45/2_1^+$ in band-1b; $41/2_2^+$ and $45/2_2^+$ in band-1a; $43/2_2^+$, $47/2_2^+$  and $51/2_1^+$  in band-2a; 
%$43/2_1^+$ and $47/2_1^+$ in band-2b are confirmed by the shell model.

\item For band-1a, the shell model results with jj44b interaction is showing reasonable agreement with the experimental data, while for JUN45 interaction results are not good for $25/2^+$--$33/2^+$ spins. Results of both interactions are not good beyond $39/2^+$ for band-2a and  band-2b.

\item The difference between shell model and experiment becomes much larger for the $47/2^+$ and $51/2^+$  states in band 2. Similar observation is also reported by the CNS calculations.

\item The order  of $11/2^+$-$13/2^+$-$15/2^+$-$19/2^+$ states in other band is correctly reproduced by the shell model.

\item The shell model result support the earlier work done using cranked Nilsson-Strutinsky calculations in Ref. \cite{R.Wadsworth} that the high-spin positive-parity states can only be formed in configurations involving three $g_{9/2}$ particles for band-1 and five particles in $g_{9/2}$ orbit in band-2. 

%%\item High-spin states in $^{67}$As nucleus come from breaking of neutron/proton pairs. The ${41/2_1}^+$ state is  a collective state because both neutrons and protons pairs are responsible to generate this state.

\item The $E-E_{rot}$ energy curve reflect the concept of configuration changes and band termination.

\item Our predicted value of quadrupole and magnetic moments might be useful to compare the future experimental data.

%\item Thus, present comprehensive study will add more informations to earlier work \cite{R.Wadsworth}.

\end{itemize}

Thus, present comprehensive study will add more information to earlier work \cite{R.Wadsworth}.
 
%\newpage

%\vspace{10cm}
\section*{Acknowledgments}
V.K. acknowledges financial support from SERB Project (File No. EEQ/2019/000084), Govt. of India. We acknowledge Prayag 5 nodes computational facility at Physics Department,
IIT-Roorkee. We would like to thank Prof. I. Ragnarsson and Prof. R. Wadsworth for useful discussions during this work.

%\newpage
%\vspace{8cm}
%%%%%%%%%%%%%%%%%%%%%%%%%%%%%%%%%%%%%%%%%%%%%%%%%%%%%%
\bibliographystyle{utphys}
  \bibliography{references}

\end{document}